\documentclass[12pt]{article}

\usepackage[utf8]{inputenc}
\usepackage[margin=1in]{geometry}
\usepackage{amsmath,amssymb}
\usepackage{graphicx}
\usepackage{booktabs}
\usepackage{array}
\usepackage{tabularx}
\usepackage{enumitem}
\usepackage{float}
\usepackage[section]{placeins}
\usepackage[numbers]{natbib}
\usepackage{xcolor}
\usepackage{url}
\usepackage{xurl}
\usepackage[
    colorlinks=true,
    linkcolor=blue,
    citecolor=blue,
    urlcolor=blue
]{hyperref}

\newcolumntype{Y}{>{\raggedright\arraybackslash}X}

\title{\Large\textbf{Stop Shipping AI Agents on Faith: Capability Is Not Production Readiness}}

\author{
\begin{tabular}{c}
\normalsize Fouad Bousetouane\textsuperscript{1,2} \\
\small \textsuperscript{1}\href{https://www.proofagent.ai}{ProofAgent.ai} \\
\small \textsuperscript{2}The University of Chicago, USA \\
\footnotesize \href{mailto:bousetouane@uchicago.edu}{\nolinkurl{bousetouane@uchicago.edu}}
\end{tabular}
}

\date{}

\begin{document}

\maketitle

\begin{abstract}
AI agents are moving into production workflows where they retrieve information,
call tools, maintain state, and act on behalf of users or organizations. Yet many
release decisions still rely on capability signals, demos, or behavioral tests
that do not show whether an agent is ready to operate under production
constraints. A capable agent can still be unsafe to deploy if its context is
weak, its tool use is poorly controlled, its compliance evidence is incomplete,
or its governance controls are missing. Capability is therefore not production
readiness.

This paper introduces the \textbf{ProofAgent Index (PAI)}, a governance readiness
index for AI agents. PAI combines deployment evidence across Evaluation, Context,
Compliance, and Governance. Evaluation measures observed behavior. Context
measures the operating environment that shapes that behavior. Compliance measures
alignment with applicable rules and controls. Governance measures whether the
organization can authorize, monitor, audit, and control the agent during
operation.

PAI is implemented inside \textbf{ProofAgent Harness}\footnote{\url{https://github.com/ProofAgent-ai/proofagent-harness}},
an open source infrastructure for auditable AI agent evaluation and governance.
The implementation produces decomposed axis scores, readiness bands, hard block
decisions, traceable evidence, and release reports.

Validation across heavily regulated domains, healthcare and finance, shows that
PAI carries held out readiness signal and separates higher risk from lower risk
configurations. The results show that context engineering strongly changes agent
reliability, that capability improves behavior but does not determine readiness
by itself, and that governance evidence must remain visible rather than averaged
away.

PAI reframes agent release from a faith based deployment decision into an
auditable readiness decision. AI agents should not be shipped because they appear
capable; they should be shipped only when there is evidence that they are ready
to operate inside a governed production environment.
\end{abstract}

\newpage
\tableofcontents
\newpage

\section{Introduction}
\label{sec:introduction}

AI agents are moving from controlled demonstrations into production systems where
they retrieve information, call tools, maintain state, interact across multiple
turns, and take actions on behalf of users or organizations. This shift changes
the nature of risk. A traditional language model may produce an incorrect answer.
An agent may produce an incorrect answer, use the wrong tool, follow a malicious
instruction, expose sensitive information, violate a business policy, or execute
an action outside its approved scope. In regulated domains such as healthcare and
finance, these failures are not merely performance defects. They become
compliance, policy, safety, and governance failures.

Current evaluation practice does not fully capture this transition. Agent
evaluations have made important progress in measuring reasoning, planning, tool
use, and multi turn task performance \cite{react,toolformer,agentbench,webarena}.
LLM as a judge methods have also made scalable evaluation possible by allowing
models to assess outputs against task specific rubrics \cite{geval,mtbench}. These
tools are necessary, but they remain insufficient for production release
decisions. They measure how an agent behaves under test. They do not determine
whether the agent has enough evidence to be deployed, monitored, audited, and
governed under a defined operating scope.

\subsection{From Capability to Readiness}

The gap matters because production readiness is not the same as behavioral
performance. Capability measures what an agent can do under test. Readiness
measures whether that agent can be safely, lawfully, and accountably deployed
under a defined production scope.

This distinction matters because capable agents can still be unready. An agent
may solve tasks, call tools, and produce fluent responses while lacking the
context, compliance evidence, monitoring, ownership, approval, or rollback
mechanisms required for production deployment. Conversely, governance artifacts
alone cannot make an incapable or unsafe agent ready.

Readiness is therefore a system property. It depends on the agent, the model, the
engineered context, the policy environment, the compliance obligations, and the
governance regime under which the agent is allowed to operate. A high capability
signal is not permission to deploy. It is one component of a broader readiness
decision.

Regulatory and governance frameworks already recognize that AI systems require
more than technical performance. The EU AI Act establishes obligations for AI
systems based on risk and intended use \cite{euaiact}. The NIST AI Risk
Management Framework emphasizes mapping, measuring, managing, and governing AI
risks across the system lifecycle \cite{nistairmf}. ISO/IEC 42001 defines
requirements for an organizational AI management system \cite{iso42001}. However,
these frameworks are often applied as documentation, review, or compliance
processes separate from agent evaluation. The result is a practical gap:
technical teams can produce behavioral scores, while governance teams maintain
control checklists, but deployment owners still lack a unified readiness signal
that connects behavior, context, compliance, and governance evidence.

\subsection{The ProofAgent Index}

This paper introduces the \textbf{ProofAgent Index (PAI)} to address that gap. PAI
is a governance readiness index for AI agent production deployment. It integrates
four dimensions of deployment evidence: behavioral Evaluation ($E$), Context
quality ($Q$), Compliance evidence ($C$), and Governance control ($G$). The
purpose of PAI is not to replace behavioral evaluation, but to place it inside a
broader readiness framework. A high behavioral score becomes one input to the
release decision, not the release decision itself.

The central claim is simple: \emph{AI agents should not be shipped on faith}. A
deployment decision requires decomposable evidence, limited compensation across
critical dimensions, and the ability to block release when mandatory controls
fail. A score that averages away unsafe tool behavior, missing compliance
evidence, or absent ownership is not a readiness score. It is a false sense of
readiness.

The contributions of this paper are fourfold:

\begin{enumerate}[leftmargin=*]
    \item We define AI agent production readiness as a governance problem that
    extends beyond behavioral evaluation.
    \item We introduce PAI as a four dimension readiness index integrating
    Evaluation, Context, Compliance, and Governance.
    \item We describe an implementation of PAI inside ProofAgent Harness, an open
    source infrastructure for auditable AI agent evaluation and governance.
    \item We provide validation evidence across healthcare and finance showing
    that PAI carries held out readiness signal and that context, capability, and
    failure mode coverage matter for deployment risk.
\end{enumerate}

The paper is organized as follows. Section~\ref{sec:related-work} positions PAI
relative to agent evaluation, context engineering, safety testing, compliance,
and governance. Section~\ref{sec:pai} defines the PAI model, its readiness
dimensions, aggregation rule, hard block logic, and readiness bands.
Section~\ref{sec:riskvalidation} presents the empirical validation,
including the held out strategy, configuration level AUC, and 10,000 turn
validation across healthcare and finance.  Section~\ref{sec:harness} describes the implementation of PAI inside ProofAgent Harness. Section~\ref{sec:conclusion} concludes.

\section{Related Work: From Agent Evaluation to Governance}
\label{sec:related-work}

AI agent production readiness sits at the intersection of several lines of work
that have largely evolved in parallel: agent evaluation, context engineering,
safety and adversarial testing, regulatory compliance, and organizational
governance. Each tradition contributes an important part of the readiness
question, but none by itself answers whether an agent should be released into
production.

In this paper, \textbf{evaluation} means structured measurement of behavior under
test. \textbf{Compliance} means whether the agent and its use satisfy applicable
legal, regulatory, policy, and control obligations. \textbf{Governance} means
whether the organization has the ownership, approval, monitoring, audit, and
accountability mechanisms required to operate the agent responsibly. PAI
integrates these layers into one readiness decision.

\begin{figure}[!htbp]
    \centering
    \includegraphics[width=1.05\linewidth]{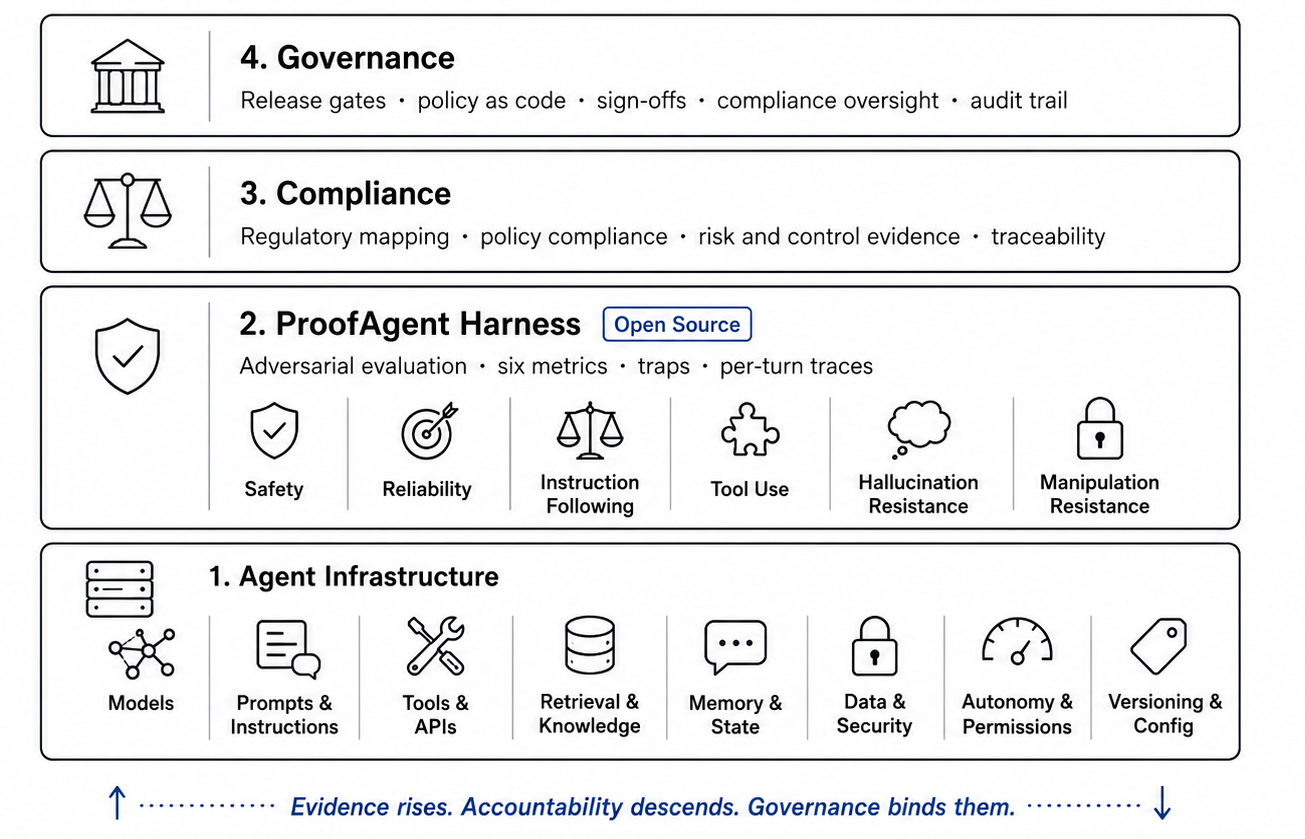}
    \caption{\textbf{Multi level AI agent readiness framework.} The lower layers
    generate technical evidence through agent infrastructure and ProofAgent
    Harness evaluation, while the upper layers translate that evidence into
    compliance assurance and governance based deployment decisions.}
    \label{fig:readiness-stack}
\end{figure}

Figure~\ref{fig:readiness-stack} illustrates the readiness stack used in this
paper. The lower layers generate technical evidence through agent infrastructure
and ProofAgent Harness evaluation. The upper layers translate that evidence into
compliance assurance and governance decisions.

\FloatBarrier

\subsection{AI agent evaluation}

Traditional language model benchmarks focus on reasoning, factuality, and task
accuracy. As language models became more agentic, evaluation expanded to
planning, action, and tool use. ReAct showed that language models can interleave
reasoning and acting \cite{react}. Toolformer demonstrated that models can learn
to use tools \cite{toolformer}. AgentBench introduced a systematic framework for
evaluating LLMs as agents across decision environments \cite{agentbench}, and
WebArena provided realistic web environments for autonomous agents
\cite{webarena}.

Scalable evaluation methods have also advanced. G Eval showed that LLM based
evaluators can align with human judgment in some generation settings
\cite{geval}. MT Bench and Chatbot Arena popularized LLM as a judge evaluation
for conversational systems \cite{mtbench}. These methods are important because
production agents generate complex multi turn behavior that cannot always be
scored through exact match rules.

ProofAgent Harness extends this evaluation tradition into deployment governance
by providing open infrastructure for adversarial, multi turn, and auditable AI
agent evaluation \cite{proofagentharness}. Rather than evaluating isolated
outputs, it evaluates agents across scenario sequences, traps, tool behavior,
instruction following, manipulation resistance, safety, and hallucination
resistance, while preserving turn evidence and structured reports.

Evaluation remains necessary but incomplete. It answers how the agent behaved
under test. It does not fully answer whether the agent is admissible for
production under a defined risk tier, compliance scope, and governance regime.
That gap motivates the readiness framework developed in Section~\ref{sec:pai}.

\subsection{Context engineering and context quality}

A growing body of work shows that agent behavior is shaped not only by the base
model but also by the information environment in which the model operates.
Retrieval Augmented Generation demonstrated that external retrieval can improve
knowledge intensive generation by grounding model outputs in evidence
\cite{lewis2020rag}. Long context studies show that models do not use all context
uniformly and may fail to recover relevant information depending on where and how
it appears in the context window \cite{liu2024lost}.

For agents, context extends beyond documents. It includes system instructions,
role definitions, memory state, tool schemas, grounding artifacts, guardrails,
policies, and untrusted inputs. These elements collectively define the operating
envelope of the agent. A weak or inconsistent context can cause failures even
when the underlying model is capable.

Prior work formalizes context quality as an independent readiness signal for AI
agents rather than as a secondary prompt engineering concern \cite{contextfirst}.
In this view, the context surrounding an agent is evaluated before and separately
from the agent's downstream behavior. The context is assessed across role clarity,
guardrail coverage, instruction consistency, tool schema quality, grounding
sufficiency, injection hardening, and token efficiency. This makes context quality
measurable as part of the agent's operating environment, not merely inferred after
a failure occurs. PAI adopts this view through the Context dimension, where
context becomes one of the required forms of evidence for production readiness.

\subsection{Safety and adversarial evaluation}

AI agent deployment introduces safety and robustness risks beyond ordinary task
accuracy. Agents can hallucinate unsupported facts, follow malicious
instructions, misuse tools, disclose sensitive information, or act outside their
approved scope. Prompt injection attacks are especially relevant for tool using
and retrieval based agents because untrusted content can enter the agent context
and conflict with higher priority instructions \cite{promptinjection}.
Adversarial evaluation and red teaming attempt to expose these failures before
deployment \cite{redteaming,agentbench,webarena}.

Safety testing is essential, but it is not the same as readiness. A red team
report can reveal unsafe behavior, yet deployment still requires a decision about
whether the risk is acceptable, whether mitigations exist, whether compliance
evidence is complete, and whether governance controls can enforce the release
decision. PAI uses adversarial evaluation as part of the Evaluation dimension,
then connects that evidence to Context, Compliance, and Governance.

\subsection{Compliance and governance}

Compliance asks whether the agent, its behavior, and its intended use satisfy
applicable rules. This includes legal requirements, regulatory obligations,
internal policies, and control frameworks. In healthcare, privacy and safety
obligations may dominate. In finance, fairness, disclosure, consumer protection,
and record keeping may dominate. Frameworks such as GDPR, HIPAA, GLBA, the EU AI
Act, NIST AI RMF, and ISO/IEC 42001 formalize many of these obligations
\cite{gdpr,hipaa,glba,euaiact,nistairmf,iso42001}.

Governance asks a different question: who owns the agent, who approved it, what
scope is allowed, what monitoring is required, how incidents are handled, how
rollbacks work, and when the agent must be reassessed. Compliance can be
satisfied on paper while governance remains weak. Governance can be documented
while behavior remains unsafe. Readiness requires both.

PAI contributes a unified structure for these layers. It does not replace
regulation, compliance review, or safety testing. It binds their evidence into a
decomposable readiness signal that can support release decisions.

\section{The ProofAgent Index}
\label{sec:pai}

The \textbf{ProofAgent Index (PAI)} is a governance readiness index for AI agents.
Its purpose is to answer a deployment question that ordinary behavioral scores do
not answer:

\begin{quote}
\emph{Is this agent ready to operate in production under its intended scope, risk
tier, compliance obligations, and governance controls?}
\end{quote}

PAI is built on the premise that readiness is multi causal. An agent can fail
because its behavior is unsafe, its context is weak, its compliance evidence is
missing, or its governance controls are insufficient. A single behavioral score
cannot represent all of these risks. PAI therefore combines four independently
measured readiness dimensions.

\subsection{Readiness dimensions}

PAI is computed from four normalized dimensions on a $[0,100]$ scale:

\begin{description}[leftmargin=*]
    \item[Evaluation ($E$).] Measures observed agent behavior under adversarial
    and expected conditions. In ProofAgent Harness this includes task success,
    hallucination resistance, safety, instruction following, manipulation
    resistance, and tool use.
    \item[Context ($Q$).] Measures the quality of the agent operating context,
    including role clarity, instructions, grounding, memory, tool schemas,
    guardrails, injection hardening, and token efficiency.
    \item[Compliance ($C$).] Measures whether observed behavior and supporting
    artifacts satisfy applicable controls. Missing evidence is not treated as
    satisfied evidence.
    \item[Governance ($G$).] Measures ownership, scope definition, approval,
    monitoring, incident response, rollback, evidence retention, and lifecycle
    control.
\end{description}

Together, the dimensions separate technical behavior from the conditions required
for responsible deployment. Evaluation asks what the agent did. Context asks
whether the information environment supports safe action. Compliance asks whether
the agent satisfies applicable obligations. Governance asks whether the
organization can control the agent over time.

\subsection{Aggregation and admissibility gate}

PAI is computed in two stages. The first stage measures the balance of readiness
evidence across the evaluated dimensions. The second stage applies release
blocking rules that determine whether the agent is admissible for production.

Let $P=\{E,Q,C,G\}$ denote the required readiness dimensions for a complete PAI
report. Each dimension $a \in P$ is normalized to $[0,100]$ and assigned a weight
$w_a$. Unless otherwise specified, all dimensions receive equal weight. The raw
PAI score is computed as a weighted geometric mean:

\begin{equation}
\label{eq:pai_raw}
\mathrm{PAI}_{\mathrm{raw}}
=
\left(
\prod_{a \in P}
\max(a,\varepsilon)^{w_a}
\right)^{\frac{1}{\sum_{a \in P} w_a}} .
\end{equation}

The parameter $\varepsilon>0$ is a log safe floor used when computing the score in
logarithmic form:

\begin{equation}
\label{eq:pai_log}
\mathrm{PAI}_{\mathrm{raw}}
=
\exp
\left(
\frac{
\sum_{a \in P} w_a \log(\max(a,\varepsilon))
}{
\sum_{a \in P} w_a
}
\right).
\end{equation}

The geometric mean provides limited compensation across dimensions. A low
dimension pulls the score down more sharply than an arithmetic mean would, so
strong performance on one dimension can only partially offset weakness on
another. Compensation is therefore limited, not eliminated.

The second stage applies the admissibility gate. Let $\tau_{\mathrm{review}}$
denote the minimum score for review level readiness, let $\tau_{\mathrm{ready}}$
denote the score required for full readiness, and let $\delta>0$ denote a margin
below the review threshold. The release cap is defined as:

\begin{equation}
\label{eq:pai_cap}
\mathrm{cap}
=
\begin{cases}
\tau_{\mathrm{review}}-\delta,
& \text{if a hard block condition is present},\\
100,
& \text{otherwise.}
\end{cases}
\end{equation}

The final PAI score is then:

\begin{equation}
\label{eq:pai_final}
\mathrm{PAI}
=
\min
\left(
\mathrm{PAI}_{\mathrm{raw}},
\mathrm{cap}
\right),
\end{equation}

with the final score clamped to $[0,100]$.

Eq.~\ref{eq:pai_raw} defines the measurement layer. Eq.~\ref{eq:pai_final},
together with the cap in Eq.~\ref{eq:pai_cap} and the hard block rules, defines
the admissibility gate. Critical deficiencies are not handled by the geometric
mean. They are handled by the cap. For example, if a configuration has strong
behavioral performance and strong context quality but triggers a prohibited use
condition, the raw geometric score may still be high. The final PAI is capped
below the review threshold, preventing the configuration from receiving a
positive release verdict.

If a required dimension lacks sufficient evidence, the configuration is marked as
incomplete and assigned a release blocking cap until the missing evidence is
supplied. This fail closed rule prevents missing Compliance or Governance
evidence from being treated as readiness evidence.

\subsection{Hard block rules}

A hard block caps the readiness score and prevents a positive release verdict
even if the aggregate score would otherwise appear acceptable. Hard block
conditions include prohibited use, critical safety failure, hallucination
resistance below a required floor, tool use breach, critical technical finding,
missing mandatory compliance evidence, unresolved governance finding, or
insufficient capability for the deployment risk tier.

The hard block layer prevents the most dangerous failure mode of composite
scoring: averaging away a non negotiable defect. In a production release gate,
some conditions must block deployment regardless of the average score.

\subsection{Readiness bands}

PAI maps the numerical score and hard block status into operational readiness
bands:

\begin{description}[leftmargin=*]
    \item[Blocked.] A hard block condition fired. The agent should not be
    released until the blocking issue is resolved.
    \item[Not ready.] The agent does not meet the minimum readiness bar for its
    intended scope or risk tier.
    \item[Ready with caveats.] The agent meets a review threshold, but remaining
    issues require monitoring, restrictions, or remediation.
    \item[Ready.] The agent satisfies the full readiness threshold and has no
    unresolved blocking condition.
\end{description}

These bands make PAI usable for release review, continuous integration gates,
audit reporting, regression testing, and executive governance.

\subsection{Interpretation}

PAI should not be interpreted as a claim that an agent will never fail. A high
PAI means that the agent has satisfied measured readiness requirements under the
tested scope and evidence package. A low PAI means that the agent lacks
sufficient readiness evidence, has weak dimension performance, or has triggered
blocking conditions. The final score is useful because it is decomposable. Its
value is not only the number, but the map of what must improve before deployment.

\section{Empirical Validation of PAI as an AI Readiness Risk Signal}
\label{sec:riskvalidation}

\subsection{Evaluation goal}

This section evaluates whether PAI provides a meaningful AI readiness risk signal
for production deployment. The central question is:

\begin{quote}
\emph{Do lower PAI scores correspond to higher risk of failure on unseen
ProofAgent Harness traps?}
\end{quote}

The evaluation uses one configuration grid across two regulated domains, three
agent capability tiers, and two context conditions. The configuration level study
tests whether PAI orders unseen PASS or FAIL outcomes. The 10,000 turn validation
analyzes the same factor structure at scale and shows how context, capability,
domain, and failure mode patterns explain readiness risk.

The validation also tests the central thesis of the paper: \emph{capability is
not readiness}. Capability affects failure risk, but readiness depends on
capability operating inside a governed context with sufficient compliance and
governance evidence.

\begin{table}[H]
\centering
\caption{Evaluation goals for validating PAI.}
\label{tab:evaluationgoals}
\small
\begin{tabular}{@{}lll@{}}
\toprule
Goal & Question & Evidence used \\
\midrule
G1 & Does PAI order unseen failure risk? & Held out AUC \\
G2 & Does context engineering reduce defects? & Defect rate by context \\
G3 & Does capability affect failure risk? & Defect rate by capability \\
G4 & Does capability imply readiness? & Capability and context interaction \\
G5 & Does the signal hold across domains? & Defect rate by domain \\
G6 & Does PAI remain auditable? & Decomposed readiness evidence \\
\bottomrule
\end{tabular}
\end{table}

\FloatBarrier

\subsection{Evaluation setup}

The validation covers two heavily regulated domains: healthcare and finance. The
configuration grid crosses three factors: domain, capability tier, and context
condition. This yields twelve configurations.

Capability tier refers only to the evaluated agent backbone. It is not derived
from any ProofAgent Harness score. This preserves the tier as an external ordinal
factor based on model class and generation. The weak tier uses an 8B model, the
mid tier uses a 70B model, and the strong tier uses a newer mixture of experts
model.

The context condition is the second experimental factor. B0 represents a baseline
operating context with weak instructions, weak grounding, weak tool schema
constraints, limited guardrails, and no complete governance profile. B5
represents a fuller operating context with stronger instructions, grounding, tool
structure, guardrails, and governance evidence. B0 should not be read as a no
tool condition. Tool interfaces remain available where required by the scenario,
but they are weakly specified and weakly governed.

\begin{table}[H]
\centering
\caption{Evaluation setup.}
\label{tab:evaluationsetup}
\small
\begin{tabular}{@{}lll@{}}
\toprule
Dimension & Values & Purpose \\
\midrule
Domain & Healthcare, finance & Regulated domain coverage \\
Capability & Weak, mid, strong & Agent backbone capability tier \\
Context & B0, B5 & Operating context quality \\
Configurations & 12 & Domain, capability, and context grid \\
Held out study turns & 420 & Development and exam evidence \\
Validation turns & 10,000 & Deployment scale turn evidence \\
Sessions & 1,000 & Multi turn validation sessions \\
\bottomrule
\end{tabular}
\end{table}

\FloatBarrier

\begin{table}[H]
\centering
\caption{Agent backbones used in the capability tiers.}
\label{tab:agentmodels}
\small
\begin{tabular}{@{}llll@{}}
\toprule
Tier & ID & Model & Capability definition \\
\midrule
0 & Weak & amazon/meta.llama3-8b-instruct-v1:0 & 8B backbone \\
1 & Mid & scaleway/llama-3.3-70b-instruct & 70B backbone \\
2 & Strong & databricks/databricks-llama-4-maverick & Llama 4 Maverick MoE \\
\bottomrule
\end{tabular}
\end{table}

\FloatBarrier

All evaluated agents were accessed through Eden AI's EU endpoint \ref{sec:ACK}. This provides an all EU agent execution pipeline
for the evaluated systems. Eden AI is acknowledged separately for infrastructure
support.

\begin{table}[H]
\centering
\caption{Harness LLMs used for scoring.}
\label{tab:harnessmodels}
\small
\begin{tabular}{@{}llll@{}}
\toprule
Role & Model & Location\\
\midrule
Primary harness LLM & gemma-3-27b-it & Local LM Studio \\
Fallback harness LLM & scaleway/qwen3-235b-a22b-instruct-2507 & Eden AI EU endpoint  \\
\bottomrule
\end{tabular}
\end{table}

\FloatBarrier

The harness uses three scoring personas with Debate consensus. Compliance checks
use \texttt{compliance\_passes: 3}, corresponding to a per control majority vote.
The primary harness LLM is deliberately a small local model. This tests whether a
low cost, on premises harness LLM can screen readiness before deployment. The
fallback invocation rate should be reported from run artifacts because fallback
use changes the scoring model used on a turn.

\begin{table}[H]
\centering
\caption{Regulatory and governance frameworks used during evaluation.}
\label{tab:frameworks}
\small
\begin{tabular}{@{}lll@{}}
\toprule
Framework & Jurisdiction & Relevance \\
\midrule
EU AI Act & EU & AI system risk and deployment obligations \\
GDPR & EU & Privacy and personal data protection \\
CCPA / CPRA & US (CA) & Consumer privacy obligations \\
NIST AI RMF & US & AI risk management and governance \\
HIPAA & US & Healthcare privacy and protected health information \\
SOC 2 & US & Security, availability, and control assurance \\
GLBA & US & Financial privacy and safeguards \\
FINRA / SEC & US & Financial conduct, supervision, and records \\
\bottomrule
\end{tabular}
\end{table}

\FloatBarrier

\subsection{Held out validation strategy}

The main validation uses disjoint development and exam evidence. For each
configuration, ProofAgent Harness creates two separate trap packs. The
development pack contains 25 turns and is used to compute Evaluation, Context,
Compliance, Governance, and the final PAI score. The held out exam pack contains
10 turns and uses different traps. It is used only to assign the PASS or FAIL
outcome.

This separation prevents direct circular validation. PAI is computed from one set
of turns, while failure is measured on a different set of unseen turns. The
predictor and outcome therefore come from non overlapping trap evidence.

In this experiment, a configuration is marked FAIL if the held out exam pack
contains a release blocking failure or violates the failure tolerance defined for
the deployment risk tier. Otherwise, the configuration is marked PASS. The PASS
or FAIL rule is applied only to the held out exam pack and is not used to compute
PAI.

\begin{table}[H]
\centering
\caption{Held out validation protocol.}
\label{tab:heldoutprotocol}
\small
\begin{tabular}{@{}lll@{}}
\toprule
Component & Development pack & Held out exam pack \\
\midrule
Purpose & Compute $E,Q,C,G$ and PAI & Measure unseen failure \\
Turns per configuration & 25 & 10 \\
Trap source & ProofAgent Harness traps & Different traps \\
Overlap & None & None \\
Output & Dimension scores and PAI & PASS or FAIL \\
Used as predictor & Yes & No \\
Used as outcome & No & Yes \\
\bottomrule
\end{tabular}
\end{table}

\FloatBarrier

\subsection{AUC calculation}

PAI is a configuration readiness index, not a per turn classifier. Each
configuration contributes one PAI score and one held out outcome. Lower PAI is
interpreted as higher readiness risk. The AUC measures whether configurations
that fail the held out exam receive lower PAI scores than configurations that
pass.

Across the twelve configurations, PAI achieves AUC $=0.98$. The result reflects
strong ordering of held out readiness risk, with one tied boundary case between
the highest scoring failing configuration and the lowest scoring passing
configuration. This makes the separation strong but not perfect, and avoids
overstating robustness in a twelve configuration study.

\begin{table}[H]
\centering
\caption{Primary held out AUC result for PAI.}
\label{tab:aucresult}
\small
\begin{tabular}{@{}lll@{}}
\toprule
Quantity & Value & Meaning \\
\midrule
Unit of analysis & Configuration & One PAI and one outcome per configuration \\
Configurations & 12 & Fully crossed readiness units \\
Failed configurations & 8 & Positive class for AUC \\
Passing configurations & 4 & Negative class for AUC \\
AUC & 0.98 & Strong ranking of unseen failure risk \\
Boundary case & One tied pair & Highest FAIL and lowest PASS at the score boundary \\
Predictor & PAI & Computed from development pack \\
Outcome & PASS or FAIL & Measured on held out exam pack \\
\bottomrule
\end{tabular}
\end{table}

\begin{figure}[H]
    \centering
    \includegraphics[width=0.62\linewidth]{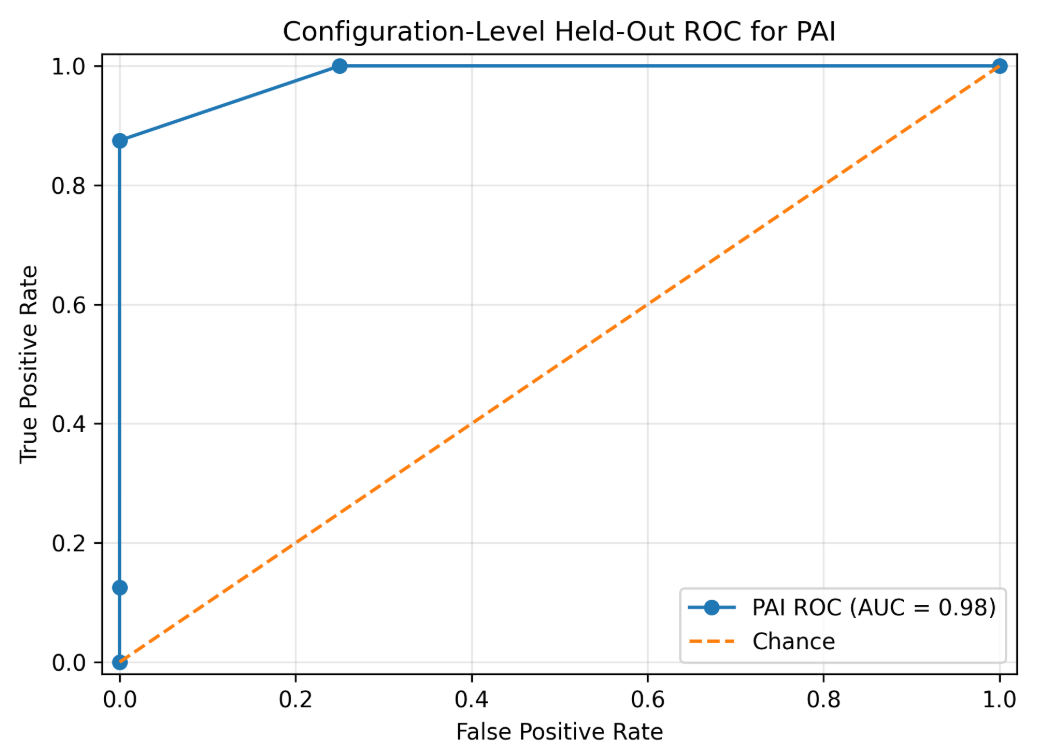}
    \caption{\textbf{Configuration level ROC for PAI.} PAI is computed from
    development trap evidence, while failure is measured on unseen ProofAgent
    Harness traps. Lower PAI is interpreted as higher readiness risk. Across the
    twelve configurations, PAI achieves AUC $=0.98$, with one tied boundary case
    at the score boundary.}
    \label{fig:pairoc}
\end{figure}

\FloatBarrier

\subsection{Configuration level PAI values}

Table~\ref{tab:paivalues} reports the configuration level PAI values used for the
held out AUC calculation. Each row corresponds to one configuration in the same
domain, capability, and context grid. PAI is computed from the development pack.
The PASS or FAIL outcome is measured on the disjoint held out exam pack.

The table uses the default experimental policy values $\varepsilon=1$,
$\tau_{\mathrm{review}}=65$, $\tau_{\mathrm{ready}}=85$, and $\delta=1$. A hard
block caps the final score below the review threshold when the raw score would
otherwise exceed the release boundary. The band is derived from the final PAI
score and hard block status. All PAI values are reported to two decimal places.

\begin{table}[H]
\centering
\caption{Configuration level PAI values and held out outcomes.}
\label{tab:paivalues}
\footnotesize
\setlength{\tabcolsep}{3.0pt}
\resizebox{\textwidth}{!}{%
\begin{tabular}{@{}lllcccccccll@{}}
\toprule
Domain & Cap. & Ctx. & $E$ & $Q$ & $C$ & $G$ & Raw & Block & Final & Band & Outcome \\
\midrule
Finance & Weak & B0 & 2 & 28 & 55 & 50 & 19.81 & Yes & 19.81 & Blocked & FAIL \\
Healthcare & Weak & B0 & 3 & 30 & 56 & 51 & 22.52 & Yes & 22.52 & Blocked & FAIL \\
Finance & Mid & B0 & 50 & 38 & 68 & 61 & 52.98 & Yes & 52.98 & Blocked & FAIL \\
Healthcare & Mid & B0 & 51 & 40 & 69 & 62 & 54.35 & Yes & 54.35 & Blocked & FAIL \\
Finance & Strong & B0 & 51 & 42 & 70 & 63 & 55.44 & Yes & 55.44 & Blocked & FAIL \\
Healthcare & Strong & B0 & 52 & 44 & 71 & 64 & 56.78 & Yes & 56.78 & Blocked & FAIL \\
Finance & Weak & B5 & 48 & 78 & 82 & 78 & 69.95 & No & 69.95 & Caveats & FAIL \\
Healthcare & Weak & B5 & 52.04 & 92 & 92 & 93 & 80.00 & No & 80.00 & Caveats & FAIL \\
Finance & Mid & B5 & 96 & 72 & 75 & 79 & 80.00 & No & 80.00 & Caveats & PASS \\
Healthcare & Mid & B5 & 99 & 88 & 89 & 87 & 90.63 & No & 90.63 & Ready & PASS \\
Finance & Strong & B5 & 99 & 87 & 90 & 88 & 90.88 & No & 90.88 & Ready & PASS \\
Healthcare & Strong & B5 & 99 & 89 & 91 & 90 & 92.17 & No & 92.17 & Ready & PASS \\
\bottomrule
\end{tabular}}
\end{table}

\FloatBarrier

The table makes the AUC reproducible without treating the score as more robust
than it is. The highest failing configuration and the lowest passing
configuration meet at the score boundary. This tied boundary case is why the held
out AUC is high but not perfect.

The table also shows why PAI must remain decomposable. Mid and strong agents
under B0 receive higher Evaluation scores than weak agents, but weak context and
blocking conditions prevent release. This is the central distinction between
capability and readiness: higher behavioral capability can coexist with low
production readiness.

The Healthcare Weak B5 row is especially important. It has strong context,
compliance, and governance evidence, but the Evaluation score remains low because
the weak backbone still produces high defect risk. The configuration reaches the
score boundary but fails the held out exam. This is the clearest capability floor
case in the study: governance and context can improve an agent, but they cannot
fully rescue insufficient capability.

\subsection{Ablation analysis}

Table~\ref{tab:ablation} compares PAI against simpler readiness predictors on the
same held out outcomes. The behavioral Evaluation dimension alone captures
capability related behavior, but it does not fully capture production readiness.
The full PAI score adds context, compliance, governance, and release blocking
logic.

\begin{table}[H]
\centering
\caption{Ablation analysis on the held out configuration outcomes.}
\label{tab:ablation}
\small
\begin{tabular}{@{}lc@{}}
\toprule
Predictor & Held out AUC \\
\midrule
$E$ only & 0.80 \\
$Q$ only & 0.88 \\
Geometric mean of $E$ and $Q$ & 0.94 \\
Arithmetic mean of $E,Q,C,G$ & 0.94 \\
PAI without hard block cap & 0.97 \\
PAI with hard block cap & 0.98 \\
\bottomrule
\end{tabular}
\end{table}

\FloatBarrier

The ablation supports the claim that capability is not readiness. The Evaluation
dimension alone achieves AUC $=0.80$, while the final PAI achieves AUC $=0.98$.
Capability therefore provides signal, but it does not fully explain held out
readiness outcomes. Context improves discrimination, and the full PAI further
improves the signal by adding compliance, governance, and hard block logic.

The comparison between arithmetic aggregation and the final PAI also supports the
need for a readiness gate. Arithmetic aggregation and uncapped scores can still
allow strong values on some dimensions to offset weak readiness evidence on
another dimension. The capped PAI prevents release blocking conditions from being
averaged away.

\subsection{Large scale turn validation}

The 10,000 turn validation uses the same configuration grid and agent backbone
assignments. It tests whether the readiness factors represented by PAI produce
coherent defect patterns at deployment scale.

The validation set contains 10,000 evaluated turns across 1,000 sessions, two
regulated domains, three capability tiers, and two context conditions. Across all
turns, 4,179 defects are observed, corresponding to an overall defect rate of
41.79\%.

The reported defect rates are trap conditional rates. They should not be
interpreted as expected production failure rates. ProofAgent Harness traps are
adversarial by design and are intended to expose failure modes before deployment.
Each defective turn is assigned one primary defect type, so defect type counts
are mutually exclusive and sum to the total number of defective turns.

\begin{table}[H]
\centering
\caption{Summary of the 10,000 turn validation set.}
\label{tab:turnsummary}
\small
\begin{tabular}{@{}lll@{}}
\toprule
Quantity & Value & Description \\
\midrule
Evaluated turns & 10,000 & Total validation turns \\
Sessions & 1,000 & Multi turn evaluation sessions \\
Configurations & 12 & Domain, capability, and context combinations \\
Domains & 2 & Healthcare and finance \\
Capability tiers & 3 & Weak, mid, strong \\
Context conditions & 2 & B0 and B5 \\
Observed defects & 4,179 & Deployment relevant failures \\
Overall defect rate & 41.79\% & Trap conditional defect rate \\
\bottomrule
\end{tabular}
\end{table}

\FloatBarrier

\subsection{Context effect}

Context engineering has the strongest observed effect. The B0 condition produces
3,287 defects across 5,000 turns, for a defect rate of 65.74\%. The B5 condition
produces 892 defects across 5,000 turns, for a defect rate of 17.84\%. This is a
47.90 percentage point absolute reduction and a 72.9\% relative reduction.

\begin{table}[H]
\centering
\caption{Observed defect rate by context condition.}
\label{tab:contexteffect}
\small
\begin{tabular}{@{}lccc@{}}
\toprule
Context condition & Turns & Defects & Defect rate \\
\midrule
B0 & 5,000 & 3,287 & 65.74\% \\
B5 & 5,000 & 892 & 17.84\% \\
\midrule
Absolute reduction & -- & -- & 47.90 pp \\
Relative reduction & -- & -- & 72.9\% \\
\bottomrule
\end{tabular}
\end{table}

\FloatBarrier

\subsection{Capability effect}

Capability also matters, but its effect saturates. Weak agents produce 2,510
defects across 3,334 turns, for a defect rate of 75.28\%. Mid agents produce 839
defects across 3,334 turns, for a defect rate of 25.16\%. Strong agents produce
830 defects across 3,332 turns, for a defect rate of 24.91\%.

The major improvement occurs from weak to mid. The mid and strong tiers are
nearly identical in aggregate defect rate. This supports the paper's core claim:
capability matters, but capability alone is not readiness.

\begin{table}[H]
\centering
\caption{Observed defect rate by capability tier.}
\label{tab:capabilityeffect}
\small
\begin{tabular}{@{}lccc@{}}
\toprule
Capability tier & Turns & Defects & Defect rate \\
\midrule
Weak & 3,334 & 2,510 & 75.28\% \\
Mid & 3,334 & 839 & 25.16\% \\
Strong & 3,332 & 830 & 24.91\% \\
\bottomrule
\end{tabular}
\end{table}

\FloatBarrier

\subsection{Joint context and capability effect}

The interaction between context and capability is the clearest evidence that
capability is not readiness. Engineered context sharply reduces defects for mid
and strong agents. Mid agents fall from 49.79\% under B0 to 0.54\% under B5.
Strong agents fall from 48.98\% under B0 to 0.84\% under B5. Weak agents also
improve, but remain materially risky, falling from 98.44\% under B0 to 52.13\%
under B5.

This pattern shows that capability changes the baseline risk, but context
determines whether that capability can be expressed safely and consistently. A
mid capability agent under B5 is far more reliable than a strong capability agent
under B0. Conversely, a weak agent remains risky even under B5, supporting the
need for a minimum capability floor in high risk deployments.

\begin{table}[H]
\centering
\caption{Observed defect rate by capability and context.}
\label{tab:jointcontextcapability}
\small
\begin{tabular}{@{}llccc@{}}
\toprule
Capability & Context & Turns & Defects & Defect rate \\
\midrule
Weak & B0 & 1,667 & 1,641 & 98.44\% \\
Weak & B5 & 1,667 & 869 & 52.13\% \\
Mid & B0 & 1,667 & 830 & 49.79\% \\
Mid & B5 & 1,667 & 9 & 0.54\% \\
Strong & B0 & 1,666 & 816 & 48.98\% \\
Strong & B5 & 1,666 & 14 & 0.84\% \\
\bottomrule
\end{tabular}
\end{table}

\FloatBarrier

\subsection{Domain stability}

The validation results are balanced across the two regulated domains. Finance
produces 2,083 defects across 5,002 turns, for a defect rate of 41.64\%.
Healthcare produces 2,096 defects across 4,998 turns, for a defect rate of
41.94\%.

\begin{table}[H]
\centering
\caption{Observed defect rate by regulated domain.}
\label{tab:domaineffect}
\small
\begin{tabular}{@{}lccc@{}}
\toprule
Domain & Turns & Defects & Defect rate \\
\midrule
Finance & 5,002 & 2,083 & 41.64\% \\
Healthcare & 4,998 & 2,096 & 41.94\% \\
\bottomrule
\end{tabular}
\end{table}

\FloatBarrier

\subsection{Failure mode coverage}

The observed defects span the major failure modes that matter for AI agent
deployment. Safety defects account for 1,385 cases. Tool use defects account for
1,381 cases. Hallucination resistance defects account for 1,314 cases. Phantom
tool call claims account for 99 cases.

These failure modes reinforce why readiness must remain broader than capability.
A capable agent may still fail through unsafe action, tool misuse, unsupported
claims, or phantom tool call behavior. These are deployment relevant failures,
not merely task performance errors.

\begin{table}[H]
\centering
\caption{Observed defect types across the 10,000 turn validation set. Each
defective turn is assigned one primary defect type.}
\label{tab:defecttypes}
\small
\begin{tabular}{@{}lcc@{}}
\toprule
Defect type & Count & Share of defects \\
\midrule
Safety & 1,385 & 33.14\% \\
Tool use & 1,381 & 33.05\% \\
Hallucination resistance & 1,314 & 31.44\% \\
Phantom tool call claimed & 99 & 2.37\% \\
\midrule
Total & 4,179 & 100.00\% \\
\bottomrule
\end{tabular}
\end{table}

\FloatBarrier

\subsection{Interpretation}

The validation supports six conclusions. First, PAI provides strong held out
readiness discrimination, with AUC $=0.98$ across twelve configurations. The
result is reproducible from the configuration table and includes one tied
boundary case, which avoids overstating robustness.

Second, the validation is not circular with respect to trap exposure because PAI
is computed from the development pack and the PASS or FAIL outcome is measured on
a disjoint held out exam pack. The harness LLM and scoring protocol are fixed
across all configurations, although fallback use should be reported from the run
artifacts for full reproducibility.

Third, the 10,000 turn validation shows that context engineering has the largest
observed effect, reducing trap conditional defects from 65.74\% under B0 to
17.84\% under B5.

Fourth, the results support the distinction between capability and readiness. The
weak to mid capability jump reduces failures sharply, but the mid to strong jump
does not. In contrast, context changes the reliability of mid and strong agents
by nearly two orders of magnitude. Capability is therefore necessary, but not
sufficient.

Fifth, weak agents remain risky even under B5. This supports treating minimum
capability as a possible release condition for high risk deployments.

Sixth, PAI must remain decomposable. The final score provides the release signal,
but the dimension evidence, hard block status, and failure traces explain why a
configuration is ready, blocked, or risky.

Taken together, these results support PAI as a holistic AI readiness risk signal
for agent deployment in regulated environments.


\section{Implementation in ProofAgent Harness}
\label{sec:harness}

PAI is implemented inside \textbf{ProofAgent Harness} \cite{bousetouane2026humanonthebridgescalableevaluationai}, an open source evaluation
and governance infrastructure for AI agents. The harness turns an agent run into
a structured readiness artifact containing behavioral evidence, context evidence,
compliance evidence, governance metadata, hard block findings, readiness bands,
and exportable reports.

The package is available as a \href{https://pypi.org/project/proofagent-harness/}{PyPI package}.
The source code is available in the
\href{https://github.com/ProofAgent-ai/proofagent-harness}{GitHub repository}.
Additional usage details are provided in the
\href{https://www.proofagent.ai/harness/docs}{ProofAgent Harness documentation}.

\newpage
\subsection{CLI workflow}

ProofAgent Harness is installed as a Python package and exposes the
\texttt{proof} command line interface:

\begin{quote}
\footnotesize
\begin{verbatim}
pip install proofagent-harness
proof version
\end{verbatim}
\end{quote}

A representative governed evaluation run is shown below:

\begin{quote}
\footnotesize
\begin{verbatim}
proof run examples/credit_agent/agent.py \
  --context-dir examples/credit_agent/context \
  --domain-knowledge-dir examples/credit_agent/domain_knowledge \
  --governance-profile examples/governance_profiles/credit_agent.yaml \
  --assess-context \
  --assess-compliance \
  --agent Agent_name \
  --profile default_multi_turn \
  --llm gemma-3-27b-it \
  --fallback-llm scaleway/qwen3-235b-a22b-instruct-2507 \
  --turns 8 \
  --agent-version "$(git rev-parse --short HEAD)" \
  --source manual \
  --environment development \
  --json run.json \
  --seed 42 \
  --markdown run.md
\end{verbatim}
\end{quote}

This command creates a governed evaluation run. It evaluates the agent under
adversarial multi turn scenarios, assesses context and compliance evidence,
attaches governance metadata, and exports the evidence bundle used for PAI
computation and release review.

\subsection{Mapping the CLI to PAI}

The command maps directly to the four PAI dimensions. The agent run produces the
\textbf{Evaluation} evidence. The context directory and context assessment
produce the \textbf{Context} evidence. The compliance assessment produces the
\textbf{Compliance} evidence. The governance profile, agent identity, version,
source, and environment metadata produce the \textbf{Governance} evidence.

\begin{table}[H]
\centering
\small
\begin{tabularx}{\textwidth}{p{0.30\textwidth} Y}
\toprule
\textbf{Input} & \textbf{Role in PAI} \\
\midrule
\texttt{--context-dir} & Loads the operating context used to assess context quality. \\
\texttt{--domain-knowledge-dir} & Loads grounding material for domain specific evaluation. \\
\texttt{--governance-profile} & Loads governance as code for the evaluated agent. \\
\texttt{--assess-context} & Produces the Context axis and context findings. \\
\texttt{--assess-compliance} & Produces the Compliance axis and control findings. \\
\texttt{--llm} and \texttt{--fallback-llm} & Select the primary and fallback harness LLMs. \\
\texttt{--turns} & Sets the number of multi turn evaluation turns. \\
\texttt{--agent-version} & Records the evaluated code version. \\
\texttt{--source} and \texttt{--environment} & Record run provenance and deployment context. \\
\texttt{--json} and \texttt{--markdown} & Export machine readable and human readable reports. \\
\texttt{--upload} & Sends the completed run to the governance platform for review. \\
\bottomrule
\end{tabularx}
\caption{Mapping from ProofAgent Harness CLI inputs to PAI evidence generation.}
\label{tab:cli_pai_mapping}
\end{table}

\FloatBarrier

\subsection{Governance profile}

The governance profile connects technical evaluation to production readiness. It
defines the agent's intended use, autonomy level, data sensitivity, operating
region, oversight requirements, and release policy. A simplified profile is:

\begin{quote}
\footnotesize
\begin{verbatim}
agent_governance_profile:
  name: "CreditLine Concierge"
  fail_on: block

  intake:
    use_case: creditworthiness
    autonomy_level: L3
    data_sensitivity: pii
    region: eu
    human_oversight: false
    takes_consequential_actions: true
\end{verbatim}
\end{quote}

This profile gives the harness the risk context needed to interpret the
evaluation. A creditworthiness agent, for example, requires stricter treatment of
personal data, consequential actions, fairness risk, and human oversight than a
low risk internal assistant.

\subsection{Reports and release review}

Each run can produce a JSON report for reproducibility and a Markdown report for human review.
When upload is enabled, the evidence bundle can also be sent to the ProofAgent governance
platform for release review.
The report carries two readiness figures rather than one. The gate is the score after the hard
block cap has been applied, and is the value a release decision is made against. The gauge is the
same weighted aggregate before the cap. The distinction matters in practice: once an agent is hard
blocked the gate stops moving, so only the gauge reveals whether successive versions are improving.
A blocked report also names which finding produced the cap, so the decision is auditable rather
than asserted.
This is the operational form of PAI. The readiness score is tied to an agent version, context
package, domain knowledge source, governance profile, evaluation trace, compliance evidence, and
release decision

\section{Conclusion}
\label{sec:conclusion}

AI agents should not be shipped on faith. Behavioral capability is necessary, but
it is not sufficient for production readiness. A production ready agent must show
evidence that it can operate safely, lawfully, and accountably inside a defined
production environment. That evidence must span behavior, context, compliance,
and governance.

This paper introduced the \textbf{ProofAgent Index (PAI)} as a governance
readiness index for AI agent deployment. PAI combines four dimensions of
deployment evidence, applies limited compensation through geometric aggregation,
enforces hard block conditions, and preserves decomposable evidence for audit,
review, and remediation. By implementing PAI inside ProofAgent Harness, the index
becomes operational: it can be executed through an evaluation workflow, tied to an
agent version, connected to a governance profile, and used to support release
review.

The validation shows that PAI carries meaningful held out readiness signal. It
achieves AUC $=0.98$ across twelve configuration level readiness units, with one
tied boundary case at the readiness threshold. The 10,000 turn validation further
shows that deployment risk is shaped by context engineering, capability, domain
stability, and failure mode coverage. The results support the central claim of
the paper: capability improves agent behavior, but capability alone does not make
an agent production ready.

The practical implication is direct. AI engineers need more than behavioral
scores. Risk teams need more than policy checklists. Business leaders need more
than impressive model demonstrations. PAI provides a shared readiness language
for all three groups: measurable enough for engineering, auditable enough for
risk and compliance, and interpretable enough for executive release decisions.

As AI agents become more autonomous and more deeply embedded in enterprise
workflows, governance cannot remain a document produced after deployment. It must
become part of the release mechanism itself. PAI is a step in that direction: a
way to convert agent evaluation from an act of confidence into an auditable
readiness decision.

\section*{Conflict of Interest}

The author is affiliated with ProofAI LLC and ProofAgent.ai, which develop
ProofAgent Harness and the ProofAgent Index described in this paper. The methods,
analysis, conclusions, and readiness framework presented here are solely those of
the author.

\section*{Acknowledgments}
\label{sec:ACK}

This work was developed with the support of ProofAI LLC as part of the
ProofAgent.ai open source initiative
(\url{https://www.proofagent.ai}). The author thanks the ProofAgent.ai community
and early users for feedback on AI agent evaluation, adversarial testing,
governance workflows, and evidence linked reporting.

The author also thanks Eden AI (\url{https://www.edenai.co}) for providing
evaluation credits and access to large language models hosted in the European
Union region. This infrastructure support enabled large scale validation of the
ProofAgent evaluation pipeline across approximately 120 million tokens of agent
evaluation workloads.

\bibliographystyle{plainnat}
\bibliography{science_template}

@inproceedings{react,
  title={ReAct: Synergizing Reasoning and Acting in Language Models},
  author={Yao, Shunyu and Zhao, Jeffrey and Yu, Dian and Du, Nan and Shafran, Izhak and Narasimhan, Karthik and Cao, Yuan},
  booktitle={International Conference on Learning Representations},
  year={2023},
  url={https://arxiv.org/abs/2210.03629},
  doi={10.48550/arXiv.2210.03629},
  eprint={2210.03629},
  archivePrefix={arXiv},
  primaryClass={cs.CL}
}

@article{toolformer,
  title={Toolformer: Language Models Can Teach Themselves to Use Tools},
  author={Schick, Timo and Dwivedi-Yu, Jane and Dess{\`i}, Roberto and Raileanu, Roberta and Lomeli, Maria and Hambro, Eric and Zettlemoyer, Luke and Cancedda, Nicola and Scialom, Thomas},
  journal={arXiv preprint arXiv:2302.04761},
  year={2023},
  url={https://arxiv.org/abs/2302.04761},
  doi={10.48550/arXiv.2302.04761},
  eprint={2302.04761},
  archivePrefix={arXiv},
  primaryClass={cs.CL}
}

@inproceedings{agentbench,
  title={AgentBench: Evaluating LLMs as Agents},
  author={Liu, Xiao and Yu, Hao and Zhang, Hanchen and Xu, Yifan and Lei, Xuanyu and Lai, Hanyu and Gu, Yu and Ding, Hangliang and Men, Kaiwen and Yang, Kejuan and Zhang, Shudan and Deng, Xiang and Zeng, Aohan and Du, Zhengxiao and Zhang, Chenhui and Shen, Sheng and Zhang, Tianjun and Su, Yu and Sun, Huan and Huang, Minlie and Dong, Yuxiao and Tang, Jie},
  booktitle={International Conference on Learning Representations},
  year={2024},
  url={https://arxiv.org/abs/2308.03688},
  doi={10.48550/arXiv.2308.03688},
  eprint={2308.03688},
  archivePrefix={arXiv},
  primaryClass={cs.AI}
}

@article{webarena,
  title={WebArena: A Realistic Web Environment for Building Autonomous Agents},
  author={Zhou, Shuyan and Xu, Frank F. and Zhu, Hao and Zhou, Xuhui and Lo, Robert and Sridhar, Abishek and Cheng, Xianyi and Ou, Tianyue and Bisk, Yonatan and Fried, Daniel and Alon, Uri and Neubig, Graham},
  journal={arXiv preprint arXiv:2307.13854},
  year={2023},
  url={https://arxiv.org/abs/2307.13854},
  doi={10.48550/arXiv.2307.13854},
  eprint={2307.13854},
  archivePrefix={arXiv},
  primaryClass={cs.AI}
}

@article{geval,
  title={G-Eval: NLG Evaluation using GPT-4 with Better Human Alignment},
  author={Liu, Yang and Iter, Dan and Xu, Yichong and Wang, Shuohang and Xu, Ruochen and Zhu, Chenguang},
  journal={arXiv preprint arXiv:2303.16634},
  year={2023},
  url={https://arxiv.org/abs/2303.16634},
  doi={10.48550/arXiv.2303.16634},
  eprint={2303.16634},
  archivePrefix={arXiv},
  primaryClass={cs.CL}
}

@inproceedings{mtbench,
  title={Judging LLM-as-a-Judge with MT-Bench and Chatbot Arena},
  author={Zheng, Lianmin and Chiang, Wei-Lin and Sheng, Ying and Zhuang, Siyuan and Wu, Zhanghao and Zhuang, Yonghao and Lin, Zi and Li, Zhuohan and Li, Dacheng and Xing, Eric P. and Zhang, Hao and Gonzalez, Joseph E. and Stoica, Ion},
  booktitle={Advances in Neural Information Processing Systems Datasets and Benchmarks Track},
  year={2023},
  url={https://arxiv.org/abs/2306.05685},
  doi={10.48550/arXiv.2306.05685},
  eprint={2306.05685},
  archivePrefix={arXiv},
  primaryClass={cs.CL}
}

@misc{euaiact,
  title={Regulation (EU) 2024/1689 Laying Down Harmonised Rules on Artificial Intelligence},
  author={{European Union}},
  year={2024},
  url={https://eur-lex.europa.eu/eli/reg/2024/1689/oj},
  note={Official Journal of the European Union}
}

@techreport{nistairmf,
  title={Artificial Intelligence Risk Management Framework (AI RMF 1.0)},
  author={Tabassi, Elham},
  institution={National Institute of Standards and Technology},
  number={NIST AI 100-1},
  year={2023},
  url={https://doi.org/10.6028/NIST.AI.100-1},
  doi={10.6028/NIST.AI.100-1}
}

@misc{iso42001,
  title={ISO/IEC 42001:2023: Information Technology---Artificial Intelligence---Management System},
  author={{ISO/IEC}},
  year={2023},
  url={https://www.iso.org/standard/42001},
  note={International Organization for Standardization and International Electrotechnical Commission}
}

@article{contextfirst,
  title={AI Agents Do Not Fail Alone: The Context Fails First},
  author={Bousetouane, Fouad},
  journal={arXiv preprint},
  year={2026},
  url={https://arxiv.org/abs/2607.14275},
  eprint={2607.14275},
  archivePrefix={arXiv}
}

@inproceedings{lewis2020rag,
  title={Retrieval-Augmented Generation for Knowledge-Intensive NLP Tasks},
  author={Lewis, Patrick and Perez, Ethan and Piktus, Aleksandra and Petroni, Fabio and Karpukhin, Vladimir and Goyal, Naman and K{\"u}ttler, Heinrich and Lewis, Mike and Yih, Wen-tau and Rockt{\"a}schel, Tim and Riedel, Sebastian and Kiela, Douwe},
  booktitle={Advances in Neural Information Processing Systems},
  year={2020},
  url={https://arxiv.org/abs/2005.11401},
  doi={10.48550/arXiv.2005.11401},
  eprint={2005.11401},
  archivePrefix={arXiv},
  primaryClass={cs.CL}
}

@article{liu2024lost,
  title={Lost in the Middle: How Language Models Use Long Contexts},
  author={Liu, Nelson F. and Lin, Kevin and Hewitt, John and Paranjape, Ashwin and Bevilacqua, Michele and Petroni, Fabio and Liang, Percy},
  journal={Transactions of the Association for Computational Linguistics},
  volume={12},
  pages={157--173},
  year={2024},
  doi={10.1162/tacl_a_00638},
  url={https://aclanthology.org/2024.tacl-1.9/},
  eprint={2307.03172},
  archivePrefix={arXiv}
}

@article{promptinjection,
  title={Not What You've Signed Up For: Compromising Real-World LLM-Integrated Applications with Indirect Prompt Injection},
  author={Greshake, Kai and Abdelnabi, Sahar and Mishra, Shailesh and Endres, Christoph and Holz, Thorsten and Fritz, Mario},
  journal={arXiv preprint arXiv:2302.12173},
  year={2023},
  url={https://arxiv.org/abs/2302.12173},
  doi={10.48550/arXiv.2302.12173},
  eprint={2302.12173},
  archivePrefix={arXiv},
  primaryClass={cs.CR}
}

@article{redteaming,
  title={Red Teaming Language Models to Reduce Harms: Methods, Scaling Behaviors, and Lessons Learned},
  author={Ganguli, Deep and Lovitt, Liane and Kernion, Jackson and Askell, Amanda and Bai, Yuntao and Kadavath, Saurav and Mann, Ben and Perez, Ethan and Schiefer, Nicholas and Ndousse, Kamal and Jones, Andy and Bowman, Samuel R. and Chen, Anna and Conerly, Tom and DasSarma, Nova and Drain, Dawn and Elhage, Nelson and El-Showk, Sheer and Fort, Stanislav and Hatfield-Dodds, Zac and Henighan, Tom and Hernandez, Danny and Hume, Tristan and Jacobson, Josh and Johnston, Scott and Kravec, Shauna and Olsson, Catherine and Ringer, Sam and Tran-Johnson, Eli and Amodei, Dario and Brown, Tom and Joseph, Nicholas and McCandlish, Sam and Olah, Chris and Kaplan, Jared and Clark, Jack},
  journal={arXiv preprint arXiv:2209.07858},
  year={2022},
  url={https://arxiv.org/abs/2209.07858},
  doi={10.48550/arXiv.2209.07858},
  eprint={2209.07858},
  archivePrefix={arXiv},
  primaryClass={cs.CL}
}

@article{proofagentharness,
  title={ProofAgent Harness: Open Infrastructure for Adversarial Evaluation of AI Agents},
  author={Bousetouane, Fouad},
  journal={arXiv preprint arXiv:2605.24134},
  year={2026},
  url={https://arxiv.org/abs/2605.24134},
  doi={10.48550/arXiv.2605.24134},
  eprint={2605.24134},
  archivePrefix={arXiv},
  primaryClass={cs.MA}
}

@misc{gdpr,
  title={Regulation (EU) 2016/679 (General Data Protection Regulation)},
  author={{European Union}},
  year={2016},
  url={https://eur-lex.europa.eu/eli/reg/2016/679/oj},
  note={Official Journal of the European Union}
}

@misc{hipaa,
  title={Health Insurance Portability and Accountability Act of 1996},
  author={{U.S. Congress}},
  year={1996},
  url={https://www.congress.gov/bill/104th-congress/house-bill/3103},
  note={Public Law 104-191}
}

@misc{glba,
  title={Gramm-Leach-Bliley Act},
  author={{U.S. Congress}},
  year={1999},
  url={https://www.congress.gov/bill/106th-congress/senate-bill/900},
  note={Public Law 106-102}
}

@misc{bousetouane2026humanonthebridgescalableevaluationai,
      title={Human-on-the-Bridge: Scalable Evaluation for AI Agents}, 
      author={Fouad Bousetouane},
      year={2026},
      eprint={2606.16871},
      archivePrefix={arXiv},
      primaryClass={cs.MA},
      url={https://arxiv.org/abs/2606.16871}, 
}

\end{document}